\documentstyle[epsfig]{aipproc}

\def\la{\mathrel{\hbox{\rlap{\hbox{\lower4pt\hbox{$\sim$}}}\hbox{$<$}}}}
\def\ga{\mathrel{\hbox{\rlap{\hbox{\lower4pt\hbox{$\sim$}}}\hbox{$>$}}}}

\def\Teff{\ifmmode{T_{\rm eff}}\else{\hbox{$T_{\rm eff}$} }\fi}
\def\Rzero{\ifmmode{R_0}\else{\hbox{$R_0$} }\fi}
\def\kms{km s$^{-1}$}

\def\56ni{$^{56}$Ni}
\def\56co{S^{56}$Co}

\def\Ch{Chandrasekhar}
\def\mni{M$_{\rm Ni}$}
\def\kms{km~s$^{-1}$}

\begin{document}
\title{Type Ia Supernovae: Toward the Standard Model?}

\author{David Branch} \address{Department of Physics and Astronomy,
University of Oklahoma, Norman, OK 73019, USA}

\maketitle

\begin{abstract}

In this short review I suggest that recent developments support the
conjecture that Type~Ia supernovae (SNe~Ia) are the complete
disruptions of Chandrasekhar--mass carbon--oxygen white dwarfs in
single--degenerate binary systems.  The causes of the observational
diversity of SNe~Ia within the context of this standard model, and the
implications of the model for young remnants of SNe~Ia, are briefly
discussed.

\end{abstract}

\section*{Introduction}

The intense current interest in Type~Ia supernovae (SNe~Ia) is mainly
due to their value as distance indicators for cosmology.  At a
conference on young supernova remnants, however, we're more concerned
with what SNe~Ia eject, and what, if anything, they leave behind.
Because what they leave behind depends on where they come from, we're
also concerned with the nature of the progenitors of SNe~Ia.  Those
who are interested in SNe~Ia as distance indicators are concerned with
these issues too.

We have a pretty good idea of what a typical SN~Ia ejects.  The
composition structure of the venerable carbon--deflagration model W7
[1] has passed the test of time as a first approximation to what is
ejected: a Chandrasekhar mass, half in the form of low--velocity
iron--peak isotopes (initially mostly radioactive $^{56}$Ni); some
high--velocity unburned carbon and oxygen; and plenty of
intermediate--mass elements such as silicon, sulfur, and calcium in
between, around 10,000~\kms.  This model, and others that are not too
unlike it, give a good account of the spectra of typical SNe~Ia [2].

W7 is a specific example of what can be considered to be the standard
model for SNe~Ia: Chandrasekhar mass ejection following the ignition
of carbon at or near the center of a carbon--oxygen (C--O) white dwarf
that has been accreting matter from a nondegenerate companion star.
In this ``single--degenerate'' (SD) scenario the donor star survives
the explosion of the white dwarf.  The Tokyo supernova group has been
advocating the standard model for years (see [3] for their most recent
review), and recently some others have been moving, more cautiously,
toward the same conclusion (see, e.g., [4] for another recent review).
I would like to climb onto this bandwagon because recent developments
make it seem to be headed in the right direction.  In this short
review the emphasis will be on these recent developments; most of the
references will be to papers that have been written within the last
few years.

\section*{Options} 

\subsection*{Sub--\Ch\ Helium Ignitors}

During the 1990s there has been some interest in sub--Chandrasekhar
off--center helium--ignitor models, in which the first nuclear
ignition occurs at or near the bottom of an accumulated helium layer
surrounding a C--O core.  Those who have calculated the
spectra and light curves of such models have found that they do not
match the observations as well as carbon--ignitor models do [5,6,7].
It is unlikely that SNe~Ia come from helium--ignitors.  If the models
change sufficiently, this conclusion can be reexamined.

\subsection*{Double Degenerates}

In the ``double--degenerate'' (DD) scenario, two white dwarfs merge
following orbital decay caused by gravitational wave radiation.  The
mass ejection can be super--Chandrasekhar (up to double--\Ch, in
principle), and no donor star is left behind.  The DD scenario has
been taken seriously, not so much because of strong evidence in its
favor, but because of perceived deficiencies in the rival SD scenario.
The main worry about the SD scenario has been that because there are
so many ways for the accreting white dwarf to lose mass (e.g., nova
explosions, common envelope formation), it may be difficult or
impossible for it to reach the \Ch\ mass.  The DD scenario provides a
natural way to assemble a super--\Ch\ mass.  Another possible problem
with the SD scenario is that some circumstellar matter is expected to
be present, while no convincing observational evidence for
circumstellar matter associated with any SN~Ia has been found.
Circumstellar matter is not expected in the DD scenario.  On the other
hand, the main worry about the DD scenario has been that most of those
who have tackled the difficult problem of modeling a white--dwarf
merger have ended up being pessimistic about getting a SN~Ia out of
it, rather than a collapse to a neutron star. The problem is that
nuclear burning begins in the outer layers and propagates inward,
converting the C--O composition to an O--Ne--Mg mixture that
eventually must collapse [8,9].  Another worry about the DD scenario
is that it may be unable to account for the rate at which SN~Ia are
observed to occur [10].

One particular SN~Ia for which a super--\Ch\ mass ejection has been
suspected is the well observed, peculiar SN~1991T.  Its light curve
was broader than the light curves of most SNe~Ia, and until the time
of maximum brightness its spectra showed strong lines of Fe~III
instead of the usual lower--excitation lines of Si~II, S~II, Ca~II,
and O~I.  Both of these properties suggest an unusually large ejected
mass of\ \ $^{56}$Ni.  If the distance to NGC~4527, the parent galaxy
of SN~1991T, is 16.4~Mpc --- the distance found from Cepheids for
NGC~4536 [11] and NGC~4496A [12] which have been thought to be
co--members with NGC~4527 of a small group of galaxies on the near
side of the Virgo cluster --- then SN~1991T appears to have been too
luminous for a \Ch--mass explosion, so it must have come from a DD
progenitor [13].  But on the basis of a simple model of the light echo
of SN~1991T from interstellar dust in NGC~4527, the maximum distance
has been estimated to be 15~Mpc [14], and recent studies of Cepheids
in NGC~4527 have given $14.1 \pm 0.8 \pm 0.8$~Mpc [15] and $13.0 \pm
0.5 \pm1.2$~Mpc [16].  (An even shorter distance of $11.3 \pm 0.4$~Mpc
has been obtained from an application of the
surface--brightness--fluctuations technique to NGC~4527 [17].)  It now
appears that as expected from its spectrum and light curve, SN~1991T
was somewhat overluminous for a SN~Ia, but not so luminous that
super--\Ch\ mass ejection and a DD progenitor are required.

It has been inferred from the spectra of SN~1991T that unburned carbon
and freshly synthesized nickel coexisted in velocity space [13], which
is not the case in any published one--dimensional hydrodynamical model
for SNe~Ia.  This characteristic of SN~1991T, although not directly
indicating a DD progenitor, has seemed to provide further evidence
that SN~1991T was different in some distinct way from normal SNe~Ia.
Recently, however, the first successful {\sl three}--dimensional
numerical simulation of a deflagration explosion of a (non--rotating)
\Ch--mass white dwarf has been carried out [18], and in the resulting
model, unburned carbon and freshly synthesized nickel do coexist in
velocity space.  It will be very interesting to see whether careful
analysis of the spectra of normal SNe~Ia reveals signs (more subtle
than in SN~1991T) that the coexistence of carbon and nickel in
velocity space is a general characteristic of SNe~Ia.

An important observational development has been the discovery of
events such as SN~1999aa that have normal spectral features near the
time of maximum brightness, but spectra that are intermediate between
those of SN~1991T (strong Fe~III) and normal SNe~Ia (strong Ca~II) a
week before maximum [19].  This seems to me [20] (but see [19]) to make
SN~1991T less likely to be physically distinct from spectroscopically
normal SNe~Ia.

All of the above results make SN~1991T begin to look more like an
event near one end of a continuum rather than like a physically
distinct phenomenon.  This is important, because if the DD scenario is
not needed for SN~1991T, then we are left with little or no evidence
that it is needed at all.  From time to time other arguments for a
range in the ejected mass of SNe~Ia have been advanced (e.g.,
[21,22]); in particular it has often been suggested that subluminous
SNe~Ia like SN~1991bg eject {\sl less} than a \Ch\ mass, but even for
such extremely peculiar events the arguments for a non--\Ch\ mass are
not compelling [23].  Given the absence of positive evidence for DD
progenitors, the lack of persuasive evidence for a range in ejected
masses among SNe~Ia, and recent developments indicating that the SD
scenario is likely to be able to produce SNe~Ia (see below), it is
tempting to think that those who have concluded that DD mergers make
neutron stars rather than SNe~Ia may have been correct.  (The recent
suggestion that a merger makes a neutron star {\sl and} a SN~Ia [24]
does not seem promising because in most cases the ejected mass would
be too much less than a \Ch\ mass.) Of course, if a single SN~Ia
proves to be too luminous for \Ch\ mass ejection (for some candidates
see [15]), this conclusion will need to be revised.

\section*{Single Degenerates}

A very attractive feature of the SD scenario is that if the accreting
white dwarf does manage to approach the \Ch\ mass, it is more likely
than a merging pair of white dwarfs to actually explode, because
unlike in the DD case burning will begin in the dense inner layers and
propagate outward.  On the issue of whether the white dwarf can get to
the \Ch\ mass, there have been important recent developments.  It has
been argued [25] that at the fairly high desired accretion rates ($
10^{-7}$ to $10^{-6}$~M$_\odot$~y$^{-1}$) the white dwarf develops a
strong fast ($\la$1000~km~s$^{-1}$) wind that stabilizes the
mass--transfer process and enables common--envelope evolution and
ruinous mass loss from the system to be avoided.  Both main--sequence
donors and red--giant donors of appropriate masses and initial
separations from the white dwarf now appear to be able to drive it to
the \Ch\ mass [26,27,28].  According to recent more detailed and
selfconsistent calculations, which take into account the evolution of
the structure of the main sequence donor, the mass transfer rate, and
the orbit, it turns out that even with less optimistic assumptions
about the role of the fast wind it is possible for the white dwarf to
reach the \Ch\ mass for a variety of initial parameters [29].  For
some systems it is not even a very close call, i.e., if the white
dwarf did not explode at 1.4~M$_\odot$ it could reach as much as
2~M$_\odot$.  Given these recent results, the tables seem to have been
turned: it now appears that SD systems with main sequence donors
should not fail to produce \Ch--mass SNe~Ia.  (And because the
accretion rate does not remain low enough long enough to build up a
dangerously massive helium layer, SD systems with main--sequence
donors do not produce sub--\Ch\ helium--ignitors.)

Observational support for the SD scenario with main--sequence donors
is provided by the existence of supersoft x--ray sources [30,31] and
recurrent novae [32,33] in which a massive white dwarf is accreting at a
suitably high rate and appears to be on its way to the \Ch\ mass.

SD systems with main sequence donors cannot produce SNe~Ia in stellar
populations much older than a Gyr. It will be interesting to see the
approach of [29] applied to SD systems with red giant donors, which
are needed in the standard model to produce the SNe~Ia that are
observed to occur in older populations.

As for the doubt about the SD scenario because of the lack of evidence
for circumstellar matter --- the existing upper limits on the amount
of circumstellar matter from optical, radio, and x--ray observations
are interesting but not sufficiently stringent to rule out the SD
scenario [34,35].  I'm not aware of any important observational
developments on this issue during the last few years.  The detection
of circumstellar matter associated with SNe~Ia should be a high
priority for observers, but if there is a fast wind from the white
dwarf, the circumstellar density will be lower than expected
previously, and detection may be very difficult.

\section*{Diversity}

If we suppose that the standard model is correct for all SNe~Ia, so
the ejected mass is always 1.4~M$_\odot$, then what causes the
observational diversity?  The primary physical variable almost
certainly is \mni, the ejected mass of $^{56}$Ni.  The higher the
value of \mni\, the higher the peak luminosity; also, because
radioactivity is responsible for heating the ejecta, the higher the
temperature and the opacity [36] and the broader the light curve
[37,38].  Thus it is likely that the observed correlation between
light--curve width and peak luminosity is mainly due to a range in
\mni.  The temperature also determines which features appear in the
early--time spectra [6,39,40].  If \mni\ is sufficiently high ($\ga
0.8$~M$_\odot$?) then Fe~III lines, instead of the usual
lower--excitation lines, become prominent.  If \mni\ is sufficiently
low ($\la 0.4$~M$_\odot$?), Ti~II lines accompany the usual lines (and
because a strong Ti~II absorption blend falls right in the middle of
the $B$ band, the $B - V$ color becomes quite red).

What causes the range in \mni?  Perhaps the most likely cause is a
range in the carbon--to--oxygen ratios of the white dwarfs, since
fusing carbon releases more energy than fusing oxygen.  The C/O ratio
is expected to be determined primarily by the main--sequence mass and
the metallicity of the progenitor of the white dwarf [41].  In the
three--dimensional modeling cited above [18], the amount of nuclear
burning does appear to depend strongly on the C/O ratio.

Not all of the observational diversity can be attributed to \mni.
Photometrically, it appears that a {\sl two--parameter} empirical
correction (using both the light--curve decline parameter, $\Delta
m_{15}$, and the maximum--light color, $B-V$) is both necessary and
(so far) sufficient to standardize the SN~Ia peak luminosities
[42,43].  Spectroscopically, a clear indication that SNe~Ia cannot be
fully described by a one--parameter sequence is that some events such
as SN~1984A have only the usual spectral features but they are
unusually broad and blueshifted, indicating that the ejecta are not
unusually hot as in the case of SN~1991T but that an unusually large
amount of mass has been ejected at high velocity ($\sim
20,000$~km~s$^{-1}$) [44,45].  This may mean that most SNe~Ia are pure
subsonic deflagrations while the high--velocity events like SN~1984A
are delayed detonations, in which the deflagration makes a transition
to a supersonic detonation with the result that the density of the
high--velocity layers is much higher than in a pure deflagration.
(For a recent review of the propagation of nuclear burning fronts in
SNe~Ia, see [46].)  Why two SNe~Ia that produce similar amounts of
\mni\ would have different modes of burning propagation remains to be
understood; or perhaps not understood --- it could be just a matter of
chance [47].  (The observed correlation between SN~Ia properties and
the characteristics of the stellar population at the SN site [47,48]
at least proves that the outcome is not {\sl entirely} unrelated to
the initial conditions.)

To summarize, it seems likely that within the SD scenario the
observational diversity can be attributed mainly to differences in
\mni, and secondarily to differences in the amount of mass ejected at
high velocity (whether both deflagrations and delayed detonations are
actually involved or not).  There are plenty of other things that
might contribute to the diversity, probably to a lesser extent: local
composition asymmetries in the ejecta of deflagrations (e.g., clumps
of iron--peak elements surrounded by intermediate--mass elements,
embedded in unburned C--O [18]); global shape asymmetries of the
ejecta in delayed detonations caused by the transition taking place at
a point (or points) rather than simultaneously everywhere on a
spherical shell [50]; a range of metallicities of the white dwarf
[51,52]; the orientation of the donor star with respect to the line of
sight of the observer [53]; effects associated with differences in the
white--dwarf rotation speeds which, owing to angular momentum acquired
during the accretion process, are expected to range up to a
significant fraction of the break--up speed [29]; and global shape
asymmetries caused by a strong magnetic field of the progenitor white
dwarf [54].

\section*{Implications for Young Supernova Remnants}

According to the standard model SN~Ia ejecta should consist of 1.4
$M_\odot$ of heavy elements from the white dwarf, accompanied by some
hydrogen--rich mass knocked off the donor star: about 0.15~$M_\odot$
from a main--sequence star and about 0.5~M$_\odot$ from a red giant
[53].  Most of this mass should be deep inside the supernova ejecta,
expanding at low velocity, less than 1000 km~s$^{-1}$ [53,55].  The
surviving donor star probably has peculiar surface abundance ratios:
no lithium, beryllium, or boron; enhanced carbon; and perhaps enhanced
iron--group elements caused by fallback of some of the
lowest--velocity supernova ejecta [29,53].  A main sequence donor will
have a space velocity (determined primarily by its pre--explosion
orbital velocity) of more than about 450~\kms\ [56] and, having been
pretty badly shaken up by its encounter with the supernova ejecta, it
will be quite overluminous, as high as 5000~L$_\odot$ [53,56] before
its swollen envelope relaxes back into equilibrium on a thermal
timescale.  A red giant donor will be moving faster than about
100~\kms\ [56], and because it will have been stripped of most but not
quite all of its envelope it will continue to look like a red giant
throughout the SNR phase [53].  Both kinds of donor stars will remain
inside their associated SNRs during the lifetime of the remnants.

Were any of the historical Galactic supernovae produced by SNe~Ia?
SNe~1006 and 1572 (Tycho's supernova) are often cited as
possibilities, but the arguments are not completely convincing.  The
heliacal risings and settings of SN~1006 have been used to make a much
improved estimate of its peak apparent magnitude, which leads to a
peak absolute visual magnitude, M$_V$, in the range $-15.9$ to $-17.4$
[57]. Normal SNe~Ia have $M_V \simeq -19.4$, so if SN~1006 was a SN~Ia
it would seem to have been a peculiar one.  If the luminosity was low
then \mni\ also was low, and not much iron ($\la 0.1$~M$_\odot$)
should be found in the remnant. (This is true whether SN~1006 was a
peculiar weak SN~Ia, a SN~Ib/c, or some other kind of peculiar event.)
If SN~1572 was a SN~Ia then it may also have been a peculiar one,
because its light curve appears to have been very fast for a SN~Ia and
its peak absolute visual magnitude has been estimated to be $-18.64
\pm 0.31$ [57], a bit subluminous for a SN~Ia and implying a reduced
iron content in the Tycho remnant, too.  The discovery in SN~1006 or
1572 of a star having the expected properties of a SN~Ia donor would
be exciting because it would confirm both that the event really was a
SN~Ia and that it was produced by a SD progenitor system; on the other
hand, establishing the {\sl absence} of a donor star would not rule
out the SD scenario if we're not completely sure that these events
were SNe~Ia. To the extent that the Balmer--dominated remnants in the
LMC are thought to have been SNe~Ia because they resemble the remnants
of SNe~1006 and 1572, then the case that the LMC events were SNe~Ia
may also not be compelling.

I am grateful to the members of the University of Oklahoma supernova
group for discussions.  This work was supported by NSF grant
AST~9986965 and NASA grant NAG5--3505.




\begin{references}

\bibitem{[]} Nomoto~K., Thielemann~F.---K., and Yokoi~K. 1984, ApJ,
286, 644 

\bibitem{[]} Lentz~E.~J., Baron~E., Branch~D., and
Hauschildt~P.~H. 2001, ApJ, in press

\bibitem{[]} Nomoto~K., Umeda~H., Kobayashi~C., Hachisu~I., Kato~M.,
and Tsujimoto~T. 2000, in Cosmic Explosions, ed. S.~S.~Holt and
W.~W.~Zhang, American Institute of Physics, in press

\bibitem{[]} Livio~M. 2000, in The Greatest Explosions 
Since the Big Bang: Supernovae and Gamma--Ray Bursts, ed. M.~Livio,
N.~Panagia, and K.~Sahu, Cambridge University Press, in press

\bibitem{[]} H\"oflich~P., Khokhlov~A., Wheeler~J.~C., Nomoto~K., and
Thielemann~F.--K. 1997, in Thermonuclear Supernovae,
ed. P.~Ruiz--Lapuente, R.~Canal, and J.~Isern, Kluwer, p.~659

\bibitem{[]} Nugent~P., Baron~E., Branch~D., Fisher~A., and
Hauschildt~P.~H. 1997, ApJ, 485, 812

\bibitem{[]} H\"oflich~P., and Khokhlov~A. 1996, ApJ, 457, 500

\bibitem{[]} Segretain~L., Chabrier~G., and Mochkovitch~R. 1997, ApJ,
481, 355

\bibitem{[]} Saio~H., and Nomoto~K. 1998, ApJ, 500, 388

\bibitem{[]} Ruiz--Lapuente~P., and Canal~R. 2001, ApJ, in press

\bibitem{[]} Saha~A., Labhardt~L., Schwengeler~H., Macchetto~F.~D.,
Panagia~N., Sandage~A., and Tammann~G.~A. 1994, ApJ, 425, 14  

\bibitem{[]} Saha~A., Sandage~A., Labhardt~L., Tammann~G.~A.,
Macchetto~F.~D., and Panagia,~N. 1996, ApJS, 107, 693

\bibitem{[]} Fisher~F., Branch~D., Hatano~K., and Baron~E. 1999,
MNRAS, 304, 67

\bibitem{[]} Sparks~W.~B., Macchetto~F., Panagia~N., Boffi~F.~R.,
Branch~D., Hazen~M., and Della Valle~M. 1999, ApJ, 523, 585

\bibitem{[]} Saha~A., Sandage~A., Thim~F., Tammann~G.~A., Labhardt~L.,
Christensen~J., Maccheto~F.~D., and Panagia~N. 2001, ApJ, in press

\bibitem{[]} Gibson~B.~K., and Stetson~P.~B. 2001, ApJ, in press

\bibitem{[]} Richtler~T., Jensen~J.~B., Tonry~J., Barris~B., and
Drenkahn~G. 2001, A\&A, in press

\bibitem{[]} Khokhlov~A. 2001, ApJ, in press

\bibitem{[]} Li~W., Filippenko~A.~V., Treffers~R.~R., Riess~A.~G.,
Hu~J., and Qiu~Y. 2001, ApJ, in press

\bibitem{[]} Branch~D. 2001, PASP, in press

\bibitem{[]} Leibundgut~B. 2001, Astr. and Astrophys. Rev., in press

\bibitem{[]} Contardo~G., Leibundgut~B., \& Vacca~W.~D. 2000, A\&A,
359, 876

\bibitem{[]} Modjaz~M., Li~W., Filippenko~A.~V., King~J.~Y.,
Leonard~D.~C., Matheson~T., and Treffers~R.~T. 2001, PASP, in press

\bibitem{[]} King~A.~R., Pringle~J.~E., and Wickramasinghe~D.~T. 2001,
MNRAS, in press

\bibitem{[]} Hachisu~I., Kato~M., and Nomoto~K. 1996, ApJ, 470, L97

\bibitem{[]} Li~X.--D., and van den Heuvel~E.~P.~J. 1997, A\&A, 322, L9

\bibitem{[]} Hachisu~I., Kato~M., Nomoto~K., and Umeda~H. 1999, ApJ,
519, 314

\bibitem{[]} Hachisu~I., Kato~M., and Nomoto~K. 1999, ApJ, 522, 487

\bibitem{[]} Langer~N., Deutschmann~A., Wellstein~S., and
H\"oflich~P. 2000, A\&A, 362, 1046

\bibitem{[]} Kahabka~P., and van den Heuvel~E.~P.~J. 1997, ARAA, 35,
69

\bibitem{[]} Kahabka~P., Puzia~T.~H., and Pietsch~W. 1999, A\&A, 347,
L43

\bibitem{[]} Hachisu~I., Kato~M., Kato~T., and Matsumoto~K. 2000, 528,
L97

\bibitem{[]} Ergma~E., Gerskevits~J., and Sarna~S. 2001, A\&A, in press

\bibitem{[]} Cumming~R., Lundqvist~P., Smith~L.~J., Pettini~M., and
King~D.~L. 1996, MNRAS, 283, 1355

\bibitem{[]} Cumming~R., and Lundqvist~P. 1996, in Advances in Stellar
Evolution, ed. R.~T.~Rood, Cambridge University Press, p. 297

\bibitem{[]} Khokhlov~A., M\"uller~E., and H\"oflich~P. 1993, A\&A,
270, 223

\bibitem{[]} Hoflich~P., Khokhlov~A., Wheeler~J.~C., Phillips~M.~M.,
Suntzeff~N.~B., and Hamuy~M. 1996, ApJ, 472, L81

\bibitem{[]} Mazzali~P., Nomoto~K., Cappellaro~E., Nakamura~T.,
Umeda~H., Iwamoto~K 2001, ApJ, in press

\bibitem{[]} Mazzali~P., Lucy~L.~B., Danziger~I.~J., Guiffes~C.,
Cappellaro~E., and Turatto~M. 1993, A\&A, 279, 447

\bibitem{[]} Hatano~K., Branch~D., Fisher~A., Millard~J., and
Baron~E. 1999, ApJS, 121, 233

\bibitem{[]} Umeda~H., Nomoto~K., Kobayashi~C., Hachisu~I., and
Kato~M. 1999, ApJ, 522, L43

\bibitem{[]} Tripp~R., and Branch~D. 1999, ApJ, 525, 209

\bibitem{[]} Parodi~B.~R., Saha~A., Sandage~A., and
Tammann~G.~A. 2000, ApJ, 540, 634

\bibitem{[]} Hatano~K., Branch~D., Lentz~E.~J., Baron~E.,
Filippenko~A.~V., and Garnavich~P.~M. 2000, ApJ, 543, L49

\bibitem{[]} Lentz~E.~J., Baron~E., Branch~D., and
Hauschildt~P.~H. 2001, ApJ, 547, in press

\bibitem{[]} Hillebrandt~W., and Niemeyer~J.~C. 2000, ARAA, 38, 191

\bibitem{[]} Sorokina~E.~I., and Blinnikov~S.~I. 2000, Astronomy
Letters, 26, 67

\bibitem{[]} Hamuy~M., Trager~S.~C., Pinto~P.~A., Phillips~M.~M.,
Schommer~R.~A., Ivanov~V., and Suntzeff~N.~B. 2000, AJ, 120, 1479

\bibitem{[]} Ivanov~V.~D., Hamuy~M., and Pinto~P.~A. 2000, ApJ, 542,
5881

\bibitem{[]} Livne~E. 1999, ApJ, 527, L97

\bibitem{[]} H\"oflich~P., Wheeler~J.~C., and Thielemann~F.--K. 1998,
ApJ, 495, 617

\bibitem{[]} Lentz~E.~J., Baron~E., Branch~D., Hauschildt~P.~H., and
Nugent~P.~E.  2000, ApJ, 530, 966

\bibitem{[]} Marietta~E., Burrows~A., and Fryxell~B. 2000, ApJS, 128,
615

\bibitem{[]} Ghezzi~C.~R., de~Gouveia Dal Pinto~E.~M., \&
Horvath~J.~E. 2001, ApJ, in press

\bibitem{[]} Chugai~N.~N. 1986, Soviet Astronomy, 30, 563

\bibitem{[]} Canal,~R., Mendez~J., and Ruiz--Lapuente~P. 2001, ApJ, in
press

\bibitem{[]} Schaefer~B.~E. 1996, ApJ, 459, 438

\end{references}
\end{document}